\documentclass[conference]{IEEEtran}
\IEEEoverridecommandlockouts

\usepackage{acronym}
\usepackage{booktabs}
\usepackage{amsmath,amssymb,amsfonts}
\usepackage{algorithmic}
\usepackage{array}
\usepackage{cite}
\usepackage{comment}
\usepackage{graphicx}
\usepackage{multirow}
\usepackage{textcomp}
\usepackage{xcolor}
\usepackage{booktabs}
\def\BibTeX{{\rm B\kern-.05em{\sc i\kern-.025em b}\kern-.08em
    T\kern-.1667em\lower.7ex\hbox{E}\kern-.125emX}}

\usepackage{hyperref}
\hypersetup{
    colorlinks=true,
    citecolor=green,
    filecolor=black,
    linkcolor=red,
    urlcolor=blue
}
\usepackage{cleveref}
\crefname{table}{Table}{Tables}
\Crefname{table}{Table}{Tables}
\crefname{figure}{Fig.}{Figs.}
\Crefname{figure}{Fig.}{Figs.}
\crefname{section}{Section}{Sections}
\Crefname{section}{Section}{Sections}

\acrodef{ai}[AI]{artificial intelligence}
\acrodef{acpf}[ACPF]{alternating current power flow}
\acrodef{acopf}[ACOPF]{alternating current optimal power flow}
\acrodef{ai}[AI]{artificial intelligence}
\acrodef{scacopf}[SC ACOPF]{security-constrained alternating current optimal power flow}
\acrodef{cpu}[CPU]{central processing unit}
\acrodefplural{cpu}[CPUs]{central processing units}
\acrodef{dcpf}[DCPF]{direct current power flow}
\acrodef{dcopf}[DCOPF]{direct current optimal power flow}
\acrodef{gpu}[GPU]{graphical processing unit}
\acrodefplural{gpu}[GPUs]{graphical processing units}
\acrodef{hpc}[HPC]{high-performance computing}
\acrodef{llm}[LLM]{large language model}
\acrodef{rag}[RAG]{retrieval augmented generation}
\acrodef{rto}[RTO]{regional transmission operator}
\acrodef{simd}[SIMD]{single-instruction multiple-data}
\acrodef{uc}[UC]{unit commitment}
\acrodefplural{uc}[UCs]{unit commitments}
\acrodef{dduc}[DDUC]{data-driven unit commitment }
\acrodef{sc}[S.C.]{South Carolina}

\begin{document}

\title{AgentiGrid: Agentic AI for Steady State and Transient Studies}

\author{
\IEEEauthorblockN{Eve Tsybina}
\IEEEauthorblockA{\textit{Oak Ridge National Laboratory} \\
Oak Ridge, TN, USA \\
tsybinae@ornl.gov}
\\
\IEEEauthorblockN{Kaleb Brunhoeber}
\IEEEauthorblockA{\textit{Oak Ridge National Laboratory} \\
Oak Ridge, TN, USA \\
brunhoeberkx@ornl.gov}
\and
\IEEEauthorblockN{Selma Grebovi\'{c}}
\IEEEauthorblockA{\textit{University of Sarajevo} \\
Sarajevo, Bosnia and Herzegovina \\
sgrebovic@etf.unsa.ba}
\\
\IEEEauthorblockN{Slaven Peles}
\IEEEauthorblockA{\textit{Oak Ridge National Laboratory} \\
Oak Ridge, TN, USA \\
peless@ornl.gov}
\and
\IEEEauthorblockN{Samim Konjicija}
\IEEEauthorblockA{\textit{University of Sarajevo} \\
Sarajevo, Bosnia and Herzegovina \\
skonjicija@etf.unsa.ba}
\and
\thanks{This manuscript has been authored by UT-Battelle, LLC under Contract No.\ DE-AC05-00OR22725 with the U.S.\ Department of Energy. The publisher acknowledges the US government license to provide public access under the DOE Public Access Plan (http://energy.gov/downloads/doe-publicaccess-plan).}
}

\maketitle

\begin{abstract}
    
As power systems worldwide struggle to meet increasing loads or higher penetration of inverter based resources, dynamic studies become of increasing importance. The rise of \ac{ai} capabilities allows to automate some of these resource intensive studies. Coupled with parallel computing, \ac{ai} is a promising technology to conduct transient and stability studies on an unprecedented scale. This study provides a proof of concept agentic \ac{ai} capable of parallel or sequential execution of phasor domain studies for a 10,000 bus system. The agent first executes a standalone simulation of a three phase symmetric fault and then executes a series of actions, by first adding a large load to a bus and then applying the fault to the same bus to stress the system. The results of the agentic simulation are consistent with human counterfactual across four vendor backends.
\end{abstract}


\section{Introduction}

Power systems are increasingly under stress in the US and EU. In the US, load growth is outpacing generation and transmission additions, causing the system reliability risks \cite{nerc2026reliability}. In the EU, the share of inverter-based  resources reached 46\% as of Q1 2026 \cite{european_renewables_2026}, which entails more complex stability conditions. And while clear regulations for power system stability exist in any jurisdiction \cite{entsoe2026,TPL001, WECCCRT4}, implementing them and properly accounting for stability risks requires repetitive and often manual work: preparing the model data, defining the event sequence, running time-domain simulations, and checking every result against the applicable criteria.

\ac{ai} technologies have enabled the development of semi-autonomous agents that are capable of automating some of this work. A review of available agentic prototypes shows that most power system agents today address this task through a similar architecture. User input is communicated to a \ac{llm} through a module, which adds function definitions and context dictionaries. The \ac{llm} then sends instructions to backend software, such as Pandapower, ParaEMT, or an inference neural network. This architecture is well documented \cite{rojas2026tutorial} and is not discussed here in detail. While early agentic studies concentrated exclusively on steady-state analysis \cite{tsybina2026agentic}, agents for dynamic studies have started appearing recently. It is difficult to provide a comprehensive literature survey because the research space is growing very fast. A search of Scopus, Google Scholar, and arXiv finds three agents that claim transient analysis capability \cite{zhang2025poweragent, shamseldein2026gridmind, wen2025xgridagent}, although only PowerAgent has publicly available code that partially supports the claims made by its developers.

PowerAgent \cite{zhang2025poweragent, mylonas2026ss, zhang2026dyn} connects \ac{ai} agents to the proprietary PowerFactory software, and can perform nontrivial \ac{dcopf} and power distribution analyses, short-circuit studies, and phasor dynamics simulations. It was tested on the IEEE 39-bus system with different \acp{llm}, including GPT-5.5, Qwen3.5, Gemini 3.1 Pro, and several Claude models. Of the three agents, it is the only one releasing code and input files. Grid-Mind \cite{shamseldein2026gridmind} reports \ac{acpf} N-1 contingency analysis for the IEEE 118-bus system with DeepSeek-V3, R1, and GPT-4o, using Pandapower as the backend. It includes ANDES-based short-circuit analysis but reports no transient results and only mentions possible use of ParaEMT. X-GridAgent \cite{wen2025xgridagent} uses GPT-5, covering \ac{acpf}, \ac{acopf}, and \ac{dcopf} N-1 contingency analyses, on systems of up to 2,000 buses for \ac{dcopf}. Its transient analysis is limited to IEC 60909 phasor short-circuit analysis without time-domain simulation.

In addition to the reviewed agents, the agent in \cite{saha2025dragent} focuses on disturbance reports, but could also be extended to study detailed transient events.


All three agents \cite{zhang2025poweragent, shamseldein2026gridmind, wen2025xgridagent} are designed to streamline and accelerate steady-state studies, transient and stability studies, or a combination thereof. This review reveals four gaps between the intended and the demonstrated dynamic capability of existing agents:

\begin{enumerate}
\item 
\emph{Time-domain stability simulation.} The transient functions of the reviewed agents are mostly short-circuit calculations. Only PowerAgent reports dynamic simulations~\cite{zhang2026dyn}.
The ``dynamic capability'' of an agent should cover rotor angle, voltage, and frequency stability~\cite{Kundur2004}. For example, an algebraic short-circuit calculation~\cite{wen2025xgridagent} does not by itself assess stability. 

\item 
\emph{Assessment against criteria.} A simulation that runs to completion does not establish stability. Apart from contingency severity ranking~\cite{zhang2026dyn}, no agent checks results against stability criteria.

\item 
\emph{Reproducibility and traceability.} Code and data are available for only one of the three agents, and even that code is not fully reproducible. Where results are reported in detail, the problem may not be on the agent side. For instance, in~\cite{zhang2026dyn}, failed runs resulted from simulator execution rather than incorrect engineering reasoning. This also raises the question of whether agent performance is bounded by the developers' familiarity with power system simulation tools.

\item 
\emph{Sequential execution.} Beyond what the agents do, one item is missed by all. All agents run simulations one at a time, although running them in parallel is the most direct way to achieve their stated goal of faster studies.

\end{enumerate}

We take existing research one step further by proposing a reproducible, high-performance, time-domain power system agent. We build on AgentiGrid~\cite{konjicija2026agentic}, originally a steady-state agent, and add functions for phasor-domain studies. The resulting agent can execute time-domain stability simulations, independently or combined with steady-state analysis. The agent code is publicly available in the ExaGO repository~\cite{peles2026exago} and verified using human-to-agent testing principles~\cite{mylonas2026ss, zhang2026dyn}. Finally, the agent executes independent scenarios in parallel. For the 10,000-bus test case, running four fault simulations in parallel yields a 1.7$\times$ speedup over sequential execution.

The remainder of this paper is organized as follows. \Cref{sec:design} presents the architecture for adding transient and stability studies to the agent. \Cref{sec:results} presents a proof of concept, first for standalone simulations and then for combined steady-state and transient stability studies. \Cref{sec:conclusions} concludes and outlines future work.

\section{Prototype architecture and study design}
\label{sec:design}

The proposed agentic \ac{ai} follows the common design principles. The functional core receives information from the frontend and interfaces it with the backend layer. The backend uses API to send the user intent to \ac{llm} models such as GPT or DeepSeek and receives back the executable structured commands. Commands are then used by the functional core to modify a case copy, send the copy to the modeling software, check the results, and update journal to send back to the \ac{llm} or the user. The loop ends when the \ac{llm} concludes that the goal was achieved, \ac{llm} concludes it cannot achieve the goal, or when iteration limit is reached (20 by default).

\begin{figure}[htb]
\centering
\includegraphics[width=\columnwidth]{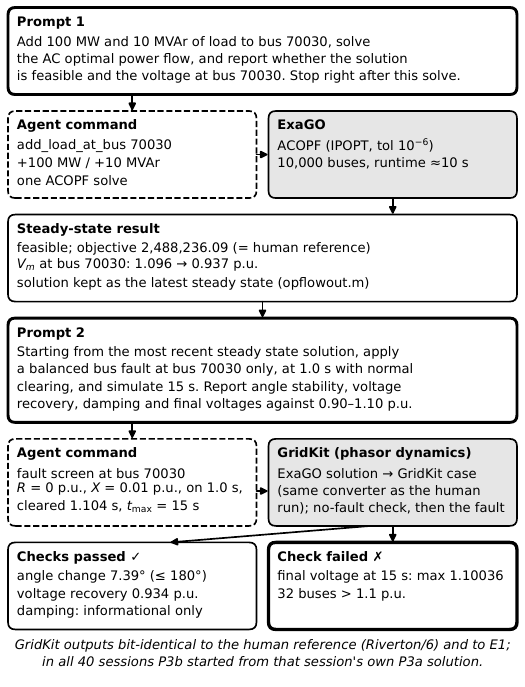}
\caption{Steady-state and transient co-simulation (E3): the second prompt starts from the agent's own \ac{acopf} solution.}
\label{fig:cosim}
\vspace{-0.5cm}
\end{figure}

The prototype architecture offers multiple advantages due to its scalability and flexibility. It allows to simulate very large systems, something that context dependent architectures that dominated the research space before 2026 cannot do. It allows for standalone or combination of different simulations. It further allows for parallel execution of multiple scenarios, something which is not done by any of the presently known \ac{ai} agents.

For the purposes of this study, the agent is expected to perform the following sequence: take a baseline case $\rightarrow$ (prompt 1) add a 100~MW, 10~Mvar load to a bus $\rightarrow$ solve the case in steady state $\rightarrow$ (prompt 2) apply a fault $\rightarrow$ solve phasor dynamics problem $\rightarrow$ report the results. To test the agentic prototype, we use an adjusted Texas A\&M 10,000 bus grid \cite{birchfield2017grid}. The agent then adds a 100 MW, 10 MVar load to the bus of interest in order to create a local stress in the system. The load approximates a large datacenter or a large industrial load with resistance prevalence.

Following \cite{mylonas2026ss, zhang2026dyn}, we first run four manual simulations to confirm that each setup is solvable and to obtain verified reference results for evaluating the agent. 
\begin{itemize}
    \item Solve the input case (objective function value 2,486,021.58, system $V_m$ range = [0.963, 1.100]).
    \item Add the 100 MW, 10 Mvar load to the baseline steady state (for example, in case of perturbed bus 70030, objective function value 2,488,236.09, system $V_m$ range = [0.937, 1.100]).
    \item Take the solved steady state case with the extra load and run 15s \cite{pjmM3} of transient analysis without any perturbations to ensure that the case can hold steady state in a dynamic simulation ($V_m$ in bus 70030 at the beginning of the simulation is 0.9373, at the end of the simulation 0.9373, largest settled $V_m$ in the system 1.1000).
    \item Add a three phase symmetric fault to the bus of interest after 1 second ($V_m$ in bus 70030 at the beginning 0.9373, at the end 0.9375, largest settled $V_m$ in the system 1.1004).

\end{itemize}
For the parallel experiment, we additionally obtained human reference simulations for the added load and faults at four buses: 70030, 13087, 10781 and 10378, built with the same system and fault parameters.

Importantly, the validation along the scheme \emph{agent + software~1} vs.\ \emph{human + software~2}, as suggested in \cite{mylonas2026ss, zhang2026dyn}, does not necessarily establish the agent's correctness. Differences between software 1 and software 2 may be significant and unrelated to the agent. We use the scheme \emph{agent + software~1} vs.\ \emph{human + software~1}, with identical scenarios, so that any discrepancy can be attributed to the agent.

All experiments ran on a single workstation (Intel Core i7-11700F, 8 cores, 64~GB RAM) with AgentiGrid from ExaGO repository (commit 07ebd2fd), built with IPOPT~3.14.20 optimization engine, and GridKit (commit 5c8b789d, built with Enzyme support). The agent was tested with four \ac{llm} backends: GLM-5.3 and DeepSeek~V4~Pro (Ollama Cloud), GPT-6 Luna (OpenAI) and Gemini~3.8 Flash (Google). 
We use another agent, Anthropic Claude, to verify prompts. We exclude Claude results from testing stages to avoid possible bias favoring that agent.
All backends used temperature~1, since GPT-6 Luna accepts no other value, and a medium reasoning effort. 

Each of the three experiments (E1: single fault, E2: parallel faults, E3: combined steady-state and transient simulation) was repeated 10 times per backend, giving 120 sessions, run one at a time. For E1 and E2, the supplied solved case was registered as the latest ExaGO steady state before each session; in E3 the agent computes that state itself. The fault is a balanced bus fault with $R=0$\,p.u. and $X=0.01$\,p.u., applied at 1.0~s and cleared after 6.25 cycles (0.104~s). Rotor-angle stability is judged by the largest change of the angle difference between any two machines relative to the pre-fault state, because the ACTIVSg10k machine angles already differ by about 108$^\circ$ in steady state. The damping ratio is reported for information only: on this multi-mode case, the peak-decay estimate depends strongly on which peaks are used. Session journals, simulation outputs, configurations and the scoring script are kept for every session.

\begin{table*}[ht]
\centering
\caption{Agent results over 30 sessions per backend (3 experiments $\times$ 10 repetitions)}
\label{tab:agent_results}
\small\setlength{\tabcolsep}{3pt}
\begin{tabular}{lcccccccc}
\hline
 & \multicolumn{3}{c}{Correct (sessions)} & \multicolumn{2}{c}{Report numbers} & \multicolumn{2}{c}{Tokens/session ($10^3$)} & Time/session (s) \\
Backend & Cmd. & Values & Verdicts & Traced$^a$ & Fully traced & Total & Output$^b$ & LLM / wall \\
\hline
GPT-6 Luna & 30/30 & 30/30 & 30/30 & 1517/1517 (100.0\%) & 30/30 & 14.2 & 1.61 & 19 / 115 \\
Gemini 3.8 Flash & 30/30 & 30/30 & 30/30 & 1676/1738 (96.4\%) & 6/30 & 18.3 & 3.47 & 28 / 125 \\
DeepSeek V4 Pro & 30/30 & 30/30 & 30/30 & 1536/1560 (98.5\%) & 19/30 & 15.9 & 3.05 & 24 / 120 \\
GLM-5.3 & 30/30 & 30/30 & 30/30 & 2144/2278 (94.1\%) & 4/30 & 23.3 & 9.84 & 72 / 168 \\
\hline
\end{tabular}
\\[2pt]\parbox{0.9\textwidth}{\footnotesize $^a$Numbers in the agent's report traced to computed values in its journal (rel.\ tol.\ $10^{-3}$). Untraced numbers are mostly correct values the journal does not store; see text. $^b$Completion plus reasoning tokens.}
\end{table*}

\section{Simulation results}
\label{sec:results}

\subsection{Standalone simulation}

We start with a simpler version, a standalone simulation of the transient modeled as a three phase symmetric fault at point of injection. The fault is applied to the bus of interest (70030, RIVERTON), located in a weakly connected part of the grid, with only two transmission lines connected to it (rate A 222.4-251.4 MVA), a native load of 54.72 MW and 35.76 MVar, and a 45 MVar capacitor inside the bus. Two types of prompts allow to simulate a fault in a single/sequential mode and parallel mode.

[Prompt 1]. Use the supplied case. Apply a balanced bus fault at bus 70030 only, at 1.0 s with normal clearing, and simulate 15 s of study. Report angle stability, voltage recovery, damping and final voltages against 0.90-1.10 pu.

[Prompt 2]. Use the supplied case. Apply a balanced bus fault to each bus in the list: 70030, 13087, 10781, 10378, at 1.0 s with normal clearing, and simulate 15 s of study. Report angle stability, voltage recovery, damping and final voltages against 0.90-1.10 pu.

The first prompt tests the regular sequential style agentic capability. The second prompt is a single instruction multiple thread parallel example: the agent interprets the user intent once and collects the results once, while the parallel core runs the four GridKit simulations simultaneously. The second prompt, therefore, is of more interest as it is a more cost efficient way of using tokens and time spent by the agent.

The results are summarized in \cref{tab:agent_results}. 
All 120 sessions issued exactly the requested commands, and every simulation output was bit-identical within each experiment and bus and matched the human reference within solver tolerance. The verdicts in the agents' reports equal the checks computed from the outputs in every session. A single fault simulation takes about 33 s, while the four simultaneous simulations of Prompt 2 finish in 77–79 s, providing backend speedup of $1.7\times$. 

\begin{figure}[htb]
\centering
\includegraphics[width=\columnwidth]{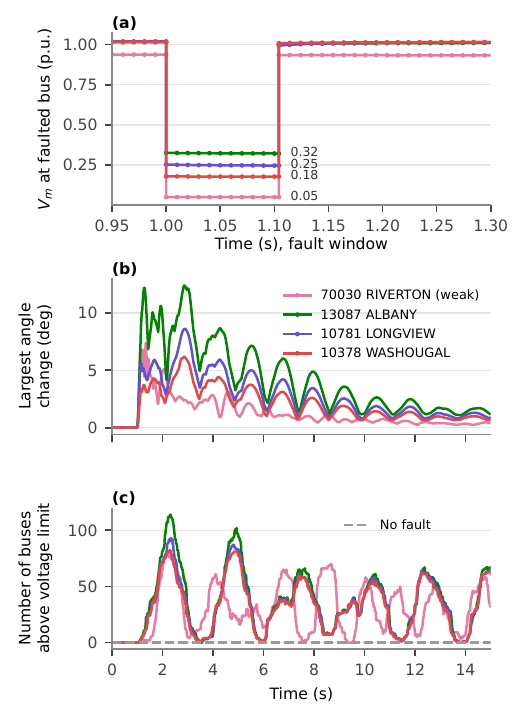}
\caption{Parallel faults (E2): (a) voltage at the faulted bus in the fault window, (b) largest change of the rotor-angle difference, (c) number of buses above the voltage limit of 1.1 p.u.}
\label{fig:e2}
\end{figure}

The agentic simulation shows that voltage at bus 70030 drops to 0.05~p.u.\ during the fault and returns to its pre-fault value after clearing, while the highest system voltage peaks at 1.108~p.u.\ and is still oscillating around the 1.10~p.u.\ limit at 15~s (\cref{fig:e2}). At this point, the agent reported ``ACTIVSg10k (faults at buses 70030) does NOT pass''.
The parallel experiment separates the weak bus from the strong ones by the shape of transient response and not necessarily the pass/fail verdict. A fault at RIVERTON (70030) depresses its voltage to 0.05~p.u., against 0.18--0.32~p.u.\ at the three strong buses, but the disturbance stays local. The largest angle change is a short 7.4$^\circ$ spike that decays to 0.3$^\circ$ by 15~s. Faults at the strong buses give shallower dips but slower, system-wide swings, up to 12.4$^\circ$ for ALBANY (13087), still 1--1.5$^\circ$ at 15~s. All four cases are angle stable and recover their voltage, and all four fail only the final-voltage check.

\subsection{Cosimulation of steady state and transient}

A more interesting case appears if the agent needs to first solve a steady state case and then apply a fault to it. This is commonly done, for instance, in connection studies. The agent receives two prompts, fed one after the other.

[Prompt 1] Add 100 MW and 10 MVAr of load to bus 70030, solve the AC optimal power flow, and report whether the solution is feasible and the voltage at bus 70030. Stop right after this solve.

[Prompt 2] Starting from the most recent steady state solution, apply a balanced bus fault at bus 70030 only, at 1.0 s with normal clearing, and simulate 15 s. Report angle stability, voltage recovery, damping and final voltages against 0.90-1.10~p.u.


In this study, the agent successfully added the load and solved the steady state, reproducing the human reference result in all 40 sessions. It further used the \ac{acopf} output of the steady state to start the simulations of the transient software GridKit. 

\begin{figure*}[th]
\centering
\includegraphics[width=\textwidth]{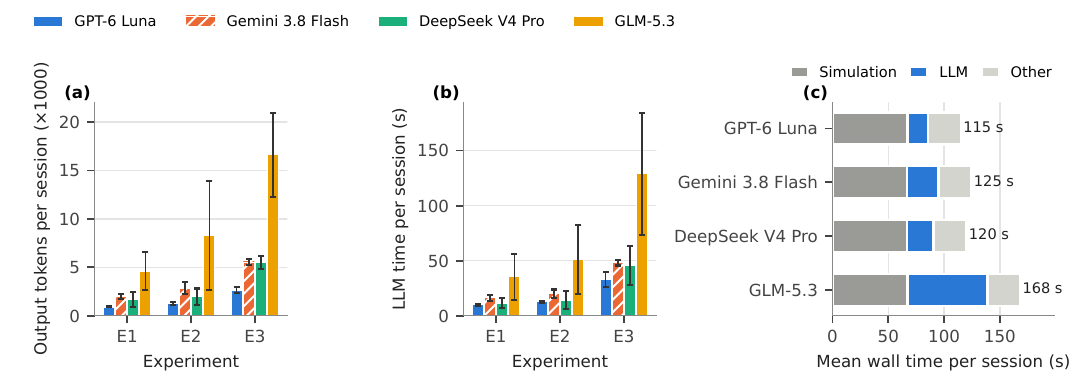}
\caption{Model comparison over 10 repetitions: (a) output tokens and (b) \ac{llm} time per session, mean and standard deviation; (c) mean wall time across all sessions split into simulation, \ac{llm} and other.}
\label{fig:models}
\end{figure*}

The result shows why it is important to automate successive steady state and transient simulations. \ac{acopf} pushes the steady state to its limits to reduce the system cost, and its solution is often recommended as the desired lowest cost dispatch. But in the \ac{acopf} steady state some of the buses end up close to the voltage limits. In the simulated 10,000-bus case, there were 25 such buses. After accounting for tolerances, 
the steady state passes, while faults at buses discussed above (70030, 13087, 10781, 10378) leave between 32 and 67 buses above the limit, by up to $4.4\cdot10^{-4}$~p.u. In addition to the necessary settling requirements, a cosimulation therefore imposes stricter starting condition requirements. This evidence shows that the true voltage magnitude limits for steady state may have to be iteratively calibrated by the agent until the system is both economically optimal and reliable.
That work is repetitive, and automating it is a promising direction where agentic \ac{ai} can significantly improve efficiency of power system operations.

\subsection{Other Results: Model comparison}

While the scope of the study is covered by the two earlier subsections, it is interesting to report several more results that are still valuable. The vendor backends differ in cost and in how closely their reports stay with the computed numbers (\cref{fig:models}). GPT-6 Luna used the fewest tokens (14.2k per session) and every number in its reports could be traced to the journal. GLM-5.3 used 1.6 times as many tokens and four times the \ac{llm} time, and had the largest share of untraced numbers (5.9\%). Most untraced numbers are correct values that the \ac{llm} does not store in the journal but evaluates on the fly. These may include the clearing time in cycles, branch loadings or generation totals.
The remainder include misattributions: in two GLM-5.3 sessions, the base-case system minimum (0.963\,p.u., bus 70173) is attributed to bus 70030, which was at 1.096\,p.u., so the voltage drop caused by the added load is understated (0.026 instead of 0.159\,p.u.). Such errors pass a number-tracing check, because every number is correct on its own.

\section{Conclusions and Future Work}
\label{sec:conclusions}

This study is an early proof of concept for a power system agentic prototype that can simulate transient to a greater detail than any of the existing prototypes, and can cosimulate steady state and transient using the same agentic core. The agent is capable of interpreting user intent, setting up a simulation, and discussing the results. The results of the agentic simulations are fully consistent with the results of a human designed simulation. Over 120 sessions with four \ac{llm} backends, the agent issued the correct commands, reproduced the human reference results within simulation tolerance, and reported verdicts consistent with the computed checks in every session. The proposed agent was operating on a relatively simple task, and further research is needed to make it create and test more sophisticated transient conditions. Further, the combined \ac{acopf} and transient simulations show that there is benefit to agentic ability to solve repeating tasks as it can find a balance between the system operating cost and reliability.


\bibliographystyle{IEEEtran}
\bibliography{references}

\end{document}